\documentclass[preprint,aps]{revtex4}

\usepackage{graphics}

\begin{document}

\title{Dirac Quasi-Normal Modes in Schwarzschild Black Hole Spacetimes}

\author{H. T. Cho}
  \email{htcho@mail.tku.edu.tw}
\affiliation{Department of Physics, Tamkang University, Tamsui,
Taipei, Taiwan, Republic of China}

\date{\today}

\begin{abstract}
We evaluate both the massless and the massive Dirac quasi-normal
mode frequencies in the Schwarzschild black hole spacetime using
the WKB approximation. For the massless case, we find that,
similar to those for the integral spin fields, the real parts of
the frequencies increase with the angular momentum number
$\kappa$, while the imaginary parts or the dampings increase with
the mode number $n$ for fixed $\kappa$. For the massive case, the
oscillation frequencies increase with the mass $m$ of the field,
while the dampings decrease. Fields with higher masses will
therefore decay more slowly.

\end{abstract}


\maketitle

\section{Introduction}

In a series of papers, Finster and collaborators
\cite{Finster1,Finster2,Finster3,Finster4,Finster5,Finster6,Finster7,Finster8,Finster9,Finster10}
studied the interaction of the Dirac field with gravity. While
they have found stable particlelike solutions in the
Einstein-Dirac-Maxwell system \cite{Finster1,Finster2}, they have
also proved the non-existence of time-periodic solutions in
various black hole spacetimes \cite{Finster4,Finster5}. This is
quite surprising because it means that Dirac particles, including
electrons and neutrinos, cannot remain on a periodic orbit around
a black hole. In \cite{Finster3}, they also showed that there are
no spherically symmetric black hole solutions in the
Einstein-Dirac-Maxwell system other than the Reissner-Nordstr\"om
one. This indicates that if a ``cloud" of Dirac particles
collapses gravitationally, the particles must eventually vanish
inside the event horizon of a black hole or escape to infinity. It
is therefore interesting to see how Dirac fields evolve in curved
background spacetimes. To understand this, Finster and
collaborators considered the massive Dirac field in the general
Newman-Kerr black hole spacetime and derived the so-called
power-law tail behavior of the field at very late time in
\cite{Finster9,Finster10}.

The evolution of wave fields around a black hole consists roughly
of three stages \cite{Frolov}. The first one is an initial wave
burst coming directly from the source and is dependent on the
initial form of the original wave field. The second one involves
the damped oscillations called the quasi-normal modes, which do
not depend on the initial values of the wave but are
characteristic of the background black hole spacetimes. The normal
mode frequencies have complex values because of radiation damping.
The last stage is the power-law tail behavior mentioned above. It
arises because of the backscattering by the long-range
gravitational field.

Here we would like to concentrate on the intermediate stage of the
evolution of the Dirac field where the quasi-normal modes
dominate. As a first step we would calculate the massive Dirac
quasi-normal mode frequencies in the Schwarzschild black hole
spacetime in this paper. The study of black hole quasi-normal
modes has a long history \cite{Kokkotas}. It started with the
black hole stability problem which concerned the evolutions of
black hole perturbations. Since these perturbations are
effectively wave fields of spins 0, 1, and 2, the evaluations of
the quasi-normal mode frequencies are mainly done for these cases.
As far as the Dirac field is concerned, we are not aware of any
work done in this direction. So we hope that our calculation can
serve to fill in this gap.

There are quite a few methods in evaluating the quasi-normal mode
frequencies. Most of them are numerical in nature. However, a
powerful analytical WKB scheme was devised by Schutz and Will
\cite{Schutz}, and was subsequently extended to higher orders in
\cite{Iyer1}. Comparing with other approaches this scheme is shown
to be accurate for the low-lying modes \cite{Iyer2}. Since we are
interested in the evolutions of the Dirac fields in the
intermediate stage, which is dominated by the low-lying modes,
this WKB scheme is good enough for our purpose.

In the next section, we consider the Dirac equation in the
Schwarzschild spacetime and its reduction into a set of
Schr\"odinger-like equations with a particular effective
potential. In Section III, we evaluate the quasi-normal
frequencies using the WKB scheme for the massless case. The
massive case is considered in Section IV. Conclusions and
discussions are presented in Section V.

\section{Dirac Equation in the Schwarzschild Spacetime}

Consider the Dirac equation in a general background spacetime
\cite{Brill},
\begin{equation}
[\gamma^{a}{e_{a}}^{\mu}(\partial_{\mu}+\Gamma_{\mu})+m]\Psi=0,
\end{equation}
where $m$ is the mass of the Dirac field, and ${e_{a}}^{\mu}$ is
the inverse of the tetrad ${e_{\mu}}^{a}$ defined by the metric
$g_{\mu\nu}$,
\begin{equation}
g_{\mu\nu}=\eta_{ab}{e_{\mu}}^{a}{e_{\nu}}^{b},
\end{equation}
with $\eta_{ab}={\rm diag}(-1,1,1,1)$ being the Minkowski metric.
$\gamma^{a}$ are the Dirac matrices
\begin{equation}
\gamma^{0}= \left(
\begin{array}{cc}
-i&0\\0&i
\end{array}\right),\ \ \
\gamma^{i}=\left(
\begin{array}{cc}
0&-i\sigma^{i}\\i\sigma^{i}&0
\end{array}\right),\ i=1,2,3,
\end{equation}
where $\sigma^{i}$ are the Pauli matrices. $\Gamma_{\mu}$ is the
spin connection given by
\begin{equation}
\Gamma_{\mu}=\frac{1}{8}[\gamma^{a},\gamma^{b}]{e_{a}}^{\nu}e_{b\nu;\mu}\
,\label{spinconnection}
\end{equation}
where
$e_{b\nu;\mu}=\partial_{\mu}e_{b\nu}-\Gamma^{\alpha}_{\mu\nu}e_{b\alpha}$
is the covariant derivative of $e_{b\nu}$ with
$\Gamma^{\alpha}_{\mu\nu}$ being the Christoffel symbols.

In the Schwarzschild spacetime,
\begin{equation}
ds^{2}=-\left(1-\frac{2M}{r}\right)dt^{2}+\left(1-\frac{2M}{r}\right)^{-1}dr^{2}
+r^{2} d\theta^{2}+r^{2}{\rm sin}^{2}\theta\ d\phi^{2},
\end{equation}
where $M$ is the mass of the black hole. Here we can take the
tetrad to be
\begin{equation}
{e_{\mu}}^{a}={\rm
diag}\left(\left(1-\frac{2M}{r}\right)^{1/2},\left(1-\frac{2M}{r}\right)^{-1/2},r,r\
{\rm sin}\theta\right).
\end{equation}
Then the spin connection defined by Eq.~(\ref{spinconnection}) is
\begin{equation}
\gamma^{a}{e_{a}}^{\mu}\Gamma_{\mu}=\gamma^{1}\left(1-
\frac{2M}{r}\right)^{1/2}\left(\frac{1}{r}+\frac{M}{2r(r-2M)}\right)+
\gamma^{2}\left(\frac{1}{2r}\right){\rm cot}\theta.
\end{equation}
Therefore, the Dirac equation becomes
\begin{eqnarray}
&&\Bigg[\gamma^{0}\left(1-\frac{2M}{r}\right)^{-1/2}\frac{\partial}{\partial
t}+\gamma^{1}\left(1-\frac{2M}{r}\right)^{1/2}\left(\frac{\partial}{\partial
r}+\frac{1}{r}+\frac{M}{2r(r-2M)}\right)\nonumber\\ &&\ \ \ \ \
+\gamma^{2}\left(\frac{1}{r}\right)\left(\frac{\partial}{\partial\theta}+
\frac{1}{2}{\rm cot}\theta\right)+\gamma^{3}\left(\frac{1}{r{\rm
sin}\theta}\right)\frac{\partial}{\partial\phi}+m\Bigg]\Psi=0.\label{diraceq}
\end{eqnarray}
The equation can be simplified by defining
\begin{equation}
\Psi=\left(1-\frac{2M}{r}\right)^{-1/4}\Phi.
\end{equation}
Then Eq.~(\ref{diraceq}) becomes
\begin{eqnarray}
&&\Bigg[\gamma^{0}\left(1-\frac{2M}{r}\right)^{-1/2}\frac{\partial}{\partial
t}+\gamma^{1}\left(1-\frac{2M}{r}\right)^{1/2}\left(\frac{\partial}{\partial
r}+\frac{1}{r}\right)\nonumber\\ &&\ \ \ \ \
+\gamma^{2}\left(\frac{1}{r}\right)\left(\frac{\partial}{\partial\theta}+
\frac{1}{2}{\rm cot}\theta\right)+\gamma^{3}\left(\frac{1}{r{\rm
sin}\theta}\right)\frac{\partial}{\partial\phi}+m\Bigg]\Phi=0,
\end{eqnarray}
which is closely related to the Dirac equation in flat spacetime
with a central potential \cite{Bjorken}. We can try the ansatz,
\begin{equation}
\Phi(t,r,\theta,\phi)= \left(\begin{array}{c}
\frac{iG^{(\pm)}(r)}{r}\varphi^{(\pm)}_{jm}(\theta,\phi)\\
\frac{F^{(\pm)}(r)}{r}\varphi^{(\mp)}_{jm}(\theta,\phi)\end{array}
\right)e^{-iEt},
\end{equation}
where for $j=l+1/2$,
\begin{equation}
\varphi^{(+)}_{jm}=\left(
\begin{array}{c}
\sqrt{\frac{l+1/2+m}{2l+1}}{Y_{l}}^{m-1/2}\\
\sqrt{\frac{l+1/2-m}{2l+1}}{Y_{l}}^{m+1/2}
\end{array}\right),
\end{equation}
and for $j=l-1/2$,
\begin{equation}
\varphi^{(-)}_{jm}=\left(
\begin{array}{c}
\sqrt{\frac{l+1/2-m}{2l+1}}{Y_{l}}^{m-1/2}\\
-\sqrt{\frac{l+1/2+m}{2l+1}}{Y_{l}}^{m+1/2}
\end{array}\right).
\end{equation}
Then the radial equations for $F^{(\pm)}$ and $G^{(\pm)}$ can be
simplified to
\begin{equation}
\frac{d}{dr_{\ast}}\left(
\begin{array}{c}
F^{(\pm)}\\G^{(\pm)}
\end{array}
\right) -\sqrt{1-\frac{2M}{r}} \left(
\begin{array}{cc}
\kappa_{(\pm)}/r&m\\ m&-\kappa_{(\pm)}/r
\end{array}
\right) \left(
\begin{array}{c}
F^{(\pm)}\\G^{(\pm)} \end{array}\right)= \left(
\begin{array}{cc}
0&-E\\E&0
\end{array}
\right) \left(
\begin{array}{c}
F^{(\pm)}\\G^{(\pm)}
\end{array}
\right),\label{FGeqn}
\end{equation}
where we have made a coordinate change
\begin{equation}
r_{\ast}=r+2M{\rm ln}\left(\frac{r}{2M}-1\right),
\end{equation}
and the constant
\begin{equation}
\kappa_{(\pm)}=\left\{
\begin{array}{cc}
-(j+1/2),&j=l+1/2\\ (j+1/2),&j=l-1/2
\end{array}
\right.
\end{equation}
Note that $\kappa_{(+)}$ and $\kappa_{(-)}$ are negative and
positive integers, respectively.

To further simplify the radial equations, we consider separately
the $(+)$ and $(-)$ cases. First, for $(+)$, we make a change of
variables \cite{Chandrasekhar}
\begin{equation}
\left(
\begin{array}{c}
\hat{F}^{(+)}\\ \hat{G}^{(+)} \end{array} \right) =\left(
\begin{array}{cc}
{\rm sin}(\theta_{(+)}/2)&{\rm cos}(\theta_{(+)}/2)\\ {\rm
cos}(\theta_{(+)}/2)&-{\rm sin}(\theta_{(+)}/2)
\end{array}\right)
\left(
\begin{array}{c}
F^{(+)}\\G^{(+)}
\end{array}
\right),
\end{equation}
where
\begin{equation}
\theta_{(+)}={\rm tan}^{-1}(mr/|\kappa_{(+)}|).
\end{equation}
Then Eq.~(\ref{FGeqn}) becomes
\begin{eqnarray}
&&\frac{d}{dr_{\ast}}\left(
\begin{array}{c}
\hat{F}^{(+)}\\ \hat{G}^{(+)}
\end{array}\right)
-\sqrt{1-\frac{2M}{r}}\sqrt{\left(\frac{\kappa_{(+)}}{r}\right)^2+m^2}
\left(
\begin{array}{cc}
1&0\\0&-1
\end{array}
\right) \left(
\begin{array}{c}
\hat{F}^{(+)}\\ \hat{G}^{(+)}
\end{array}\right) \nonumber\\ &&\ \ \ \ \
\ =-E\left(1+\frac{1}{2E}\left(1-\frac{2M}{r}\right)
\frac{m|\kappa_{(+)}|}{\kappa_{(+)}^{2}+m^{2}r^{2}}\right)
\left(\begin{array}{cc} 0&-1\\1&0 \end{array}\right) \left(
\begin{array}{c}
\hat{F}^{(+)}\\ \hat{G}^{(+)}
\end{array} \right).
\end{eqnarray}
This equation can be simplified further if we make yet another
change of variable,
\begin{equation}
\hat{r}_{\ast}=r_{\ast}+\frac{1}{2E}{\rm
tan}^{-1}\left(\frac{mr}{|\kappa_{(+)}|}\right),
\end{equation}
and then
\begin{equation}
\frac{d}{d\hat{r}_{\ast}}\left(
\begin{array}{c}
\hat{F}^{(+)}\\ \hat{G}^{(+)}
\end{array}
\right) +W_{(+)}\left(
\begin{array}{c}
-\hat{F}^{(+)}\\ \hat{G}^{(+)}
\end{array}
\right)=E\left(
\begin{array}{c}
\hat{G}^{(+)}\\ -\hat{F}^{(+)}
\end{array}
\right),
\end{equation}
where
\begin{equation}
W_{(+)}=\frac{\sqrt{1-\frac{2M}{r}}\sqrt{\left(\frac{\kappa_{(+)}}{r}\right)^{2}+m^{2}}}
{1+\frac{1}{2E}\left(1-\frac{2M}{r}\right)\left(\frac{m|\kappa_{(+)}|}
{\kappa_{(+)}^2+m^{2}r^{2}}\right)}.\label{W+}
\end{equation}
These equations for $\hat{F}^{(+)}$ and $\hat{G}^{(+)}$ can be
decoupled giving
\begin{eqnarray}
\left(-\frac{d^{2}}{d\hat{r}_{\ast}^{2}}+V_{(+)1}\right)\hat{F}^{(+)}
&=&E^{2}\hat{F}^{(+)},\\
\left(-\frac{d^{2}}{d\hat{r}_{\ast}^{2}}+V_{(+)2}\right)\hat{G}^{(+)}
&=&E^{2}\hat{G}^{(+)},
\end{eqnarray}
where
\begin{equation}
V_{(+)1,2}=\pm\frac{dW_{(+)}}{d\hat{r}_{\ast}}+W_{(+)}^{2}.
\end{equation}
For the case of $(-)$, similar procedures can be carried out
giving
\begin{eqnarray}
\left(-\frac{d^{2}}{d\hat{r}_{\ast}^{2}}+V_{(-)1}\right)\hat{F}^{(-)}
&=&E^{2}\hat{F}^{(-)},\\
\left(-\frac{d^{2}}{d\hat{r}_{\ast}^{2}}+V_{(-)2}\right)\hat{G}^{(-)}
&=&E^{2}\hat{G}^{(-)},
\end{eqnarray}
where
\begin{equation}
V_{(-)1,2}=\pm\frac{dW_{(-)}}{d\hat{r}_{\ast}}+W_{(-)}^{2},
\end{equation}
with
\begin{equation}
W_{(-)}=\frac{\sqrt{1-\frac{2M}{r}}\sqrt{\left(\frac{\kappa_{(-)}}{r}\right)^{2}+m^{2}}}
{1-\frac{1}{2E}\left(1-\frac{2M}{r}\right)\left(\frac{m\kappa_{(-)}}
{\kappa_{(-)}^2+m^{2}r^{2}}\right)}.\label{W-}
\end{equation}
From the forms of $W_{(\pm)}$ in Eqs.~(\ref{W+}) and (\ref{W-}),
we see that the $(+)$ and $(-)$ cases can be put together giving
\begin{eqnarray}
\left(-\frac{d^{2}}{d\hat{r}_{\ast}^{2}}+V_{1}\right)\hat{F}
&=&E^{2}\hat{F},\label{V1}\\
\left(-\frac{d^{2}}{d\hat{r}_{\ast}^{2}}+V_{2}\right)\hat{G}
&=&E^{2}\hat{G},\label{V2}
\end{eqnarray}
where
\begin{equation}
V_{1,2}=\pm\frac{dW}{d\hat{r}_{\ast}}+W^{2},\label{V12}
\end{equation}
with
\begin{eqnarray}
W&=&\frac{\sqrt{1-\frac{2M}{r}}\sqrt{\left(\frac{\kappa}{r}\right)^{2}+m^{2}}}
{1+\frac{1}{2E}\left(1-\frac{2M}{r}\right)\left(\frac{m\kappa}
{\kappa^{2}+m^{2}r^{2}}\right)}\nonumber\\ &=&
\frac{\Delta^{1/2}(\kappa^{2}+m^{2}r^{2})^{3/2}}{r^{2}(\kappa^{2}+m^{2}r^{2})+m\kappa
\Delta/2E}, \label{W}
\end{eqnarray}
where $\Delta=r(r-2M)$. Here $\kappa$ goes over all positive and
negative integers. Positive integers represent the $(+)$ case with
\begin{equation}
\kappa=j+\frac{1}{2}\ \ \ {\rm and}\ \ \  j=l+\frac{1}{2}
\end{equation}
while negative integers represent the $(-)$ case with
\begin{equation}
\kappa=-\left(j+\frac{1}{2}\right)\ \ \ {\rm and}\ \ \
j=l-\frac{1}{2}
\end{equation}

From the Schr\"odinger-like equations in Eqs.~(\ref{V1}) and
(\ref{V2}), we shall evaluate the corresponding quasi-normal mode
frequencies. Note that $V_{1}$ and $V_{2}$, which are related as
shown in Eq.~(\ref{V12}), are supersymmetric partners derived from
the same superpotential $W$ \cite{Cooper}. It has been shown that
potentials related in this way possess the same spectra of
quasi-normal mode frequencies \cite{Anderson}. Physically this
just indicates that Dirac particles and antiparticles have the
same quasi-normal mode spectra in the Schwarzshild black hole
spacetime which is quite reasonable here. We shall therefore
concentrate just on Eq.~(\ref{V1}) with potential $V_{1}$ in
evaluating the quasi-normal mode frequencies in the next sections.

\section{\label{section3}Quasi-normal mode frequencies for the massless Dirac field}

In this section we shall evaluate the quasi-normal frequencies for
the massless Dirac field using the WKB approximation. For the
massless case, the radial equation (Eq.~(\ref{V1})) is simplified
to
\begin{equation}
\left(-\frac{d^{2}}{dr_{\ast}^{2}}+V(r,\kappa)\right)\hat{F}=E^{2}\hat{F},
\end{equation}
where
\begin{equation}
V(r,\kappa)=\frac{|\kappa|\Delta^{1/2}}{r^{4}}[|\kappa|\Delta^{1/2}-(r-3)],
\label{m0V}
\end{equation}
$r_{\ast}=r+2\ {\rm ln}(r/2-1)$, and $\Delta=r(r-2)$. Note that we
have written $V_{1}$ as $V$ because we shall not work with
$V_{2}$, which will give the same spectrum of quasi-normal mode
frequencies, as discussed in the last section. Also we have used
the mass $M$ of the black hole as a unit of mass and length to
simplify the notation.

\begin{figure}[t]
\includegraphics{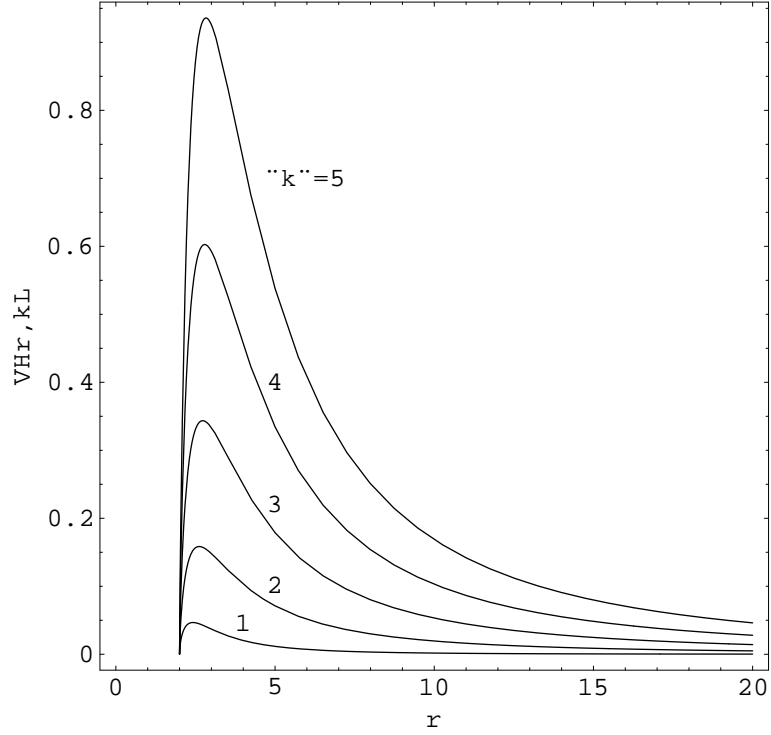}
\caption{\label{m0kappa1} Variation of the effective potential for
the massless Dirac field with $|\kappa|$.}
\end{figure}

\begin{figure}[!]
\includegraphics{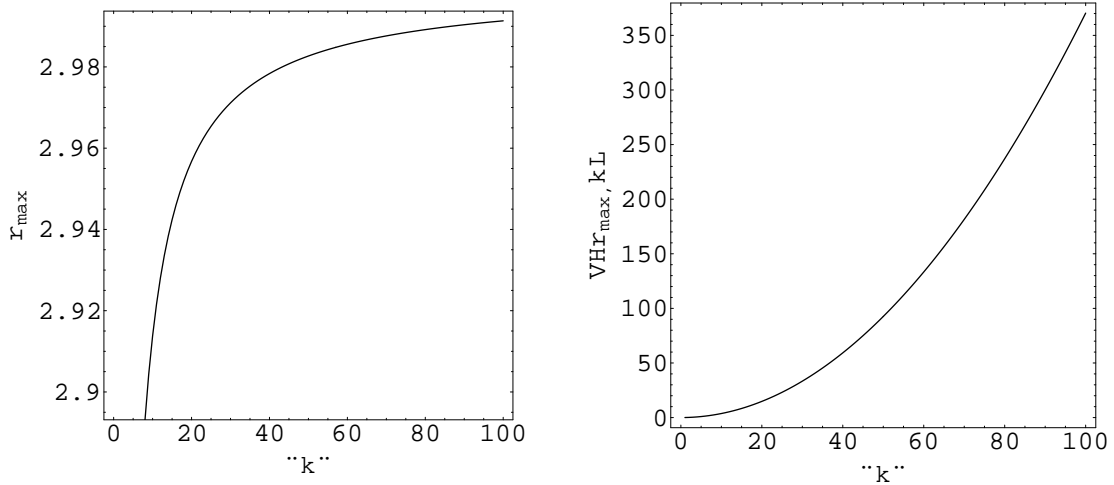}
\caption{\label{m0max1} Variation of the positions of the peaks
(left) and the peak values (right) of the effective potential for
the massless Dirac field with $|\kappa|$.}
\end{figure}

The effective potential $V(r,\kappa)$, which depends only on the
absolute value of $\kappa$, is in the form of a barrier. As shown
in Fig.~\ref{m0kappa1}, the peak of the barrier gets higher and
higher as $|\kappa|$ increases. From Fig.~\ref{m0max1}, we see
that the position of the peak
\begin{equation}
r_{max}(|\kappa|\rightarrow\infty)\rightarrow 3,
\end{equation}
and the maximum values of the potential barrier increase
indefinitely with $|\kappa|$. In fact,
\begin{equation}
V(|\kappa|\rightarrow\infty,r_{max})=\frac{\kappa^{2}}{27}
\end{equation}
as can be easily derived from Eq.~(\ref{m0V}).

\begin{table}
\caption{\label{table1}Massless Dirac quasi-normal mode
frequencies. }
\begin{ruledtabular}
\begin{tabular}{cccc}
$|\kappa|$&$n$&Re($E$)&Im($E$)\\ \hline 1 & 0 & 0.176 & -0.100
\\ 2 & 0 & 0.379 & -0.097 \\ & 1 & 0.354 & -0.299 \\ 3 & 0 &
0.574 & -0.096
\\ & 1& 0.556 & -0.293 \\ & 2 & 0.527 & -0.497 \\ 4 & 0 & 0.767 & -0.096 \\ & 1 & 0.754 & -0.291
\\ & 2 & 0.730 & -0.491 \\ & 3 & 0.700 & -0.696 \\ 5 & 0 & 0.960 & -0.096 \\ & 1 & 0.950 & -0.290
\\ & 2 & 0.930 & -0.488 \\ & 3 & 0.904 & -0.689 \\ & 4 & 0.872 & -0.894 \\
\end{tabular}
\end{ruledtabular}
\end{table}

To evaluate the quasi-normal mode frequencies, we adopt the WKB
approximation developed by Schutz, Will, and Iyer
\cite{Schutz,Iyer1,Iyer2}. This analytic method has been used
extensively in various black hole cases. Comparing with other
numerical results, this method has been found to be accurate up to
around 1\% for both the real and the imaginary parts of the
frequencies for low-lying modes with $n<l$, where $n$ is the mode
number and $l$ is the angular momentum quantum number. The formula
for the complex quasi-normal mode frequencies $E$ in the WKB
approximation, carried to third order beyond the eikonal
approximation, is given by \cite{Iyer1}
\begin{equation}
E^{2}=[V_{0}+(-2V^{''}_{0})^{1/2}\Lambda]
-i(n+\frac{1}{2})(-2V^{''}_{0})^{1/2}(1+\Omega), \label{WKBeq}
\end{equation}
where
\begin{eqnarray}
\Lambda&=&\frac{1}{(-2V^{''}_{0})^{1/2}}
\left\{\frac{1}{8}\left(\frac{V_{0}^{(4)}}{V_{0}^{''}}\right)
\left(\frac{1}{4}+\alpha^{2}\right)- \frac{1}{288}
\left(\frac{V_{0}^{'''}}{V_{0}^{''}}\right)^{2}(7+60\alpha^{2})\right\},\\
\Omega&=&\frac{1}{(-2V_{0}^{''})}\left\{\frac{5}{6912}
\left(\frac{V_{0}^{'''}}{V_{0}^{''}}\right)^{4}(77+188\alpha^{2})
-\frac{1}{384}\left(\frac{{V_{0}^{'''}}^{2}V_{0}^{(4)}}{{V_{0}^{''}}^{3}}\right)
(51+100\alpha^{2})\right.\nonumber\\ &&\ \ \
+\frac{1}{2304}\left(\frac{V_{0}^{(4)}}{V_{0}^{''}}\right)^{2}(67+68\alpha^{2})
+\frac{1}{288}\left(\frac{V_{0}^{'''}V_{0}^{(5)}}{{V_{0}^{''}}^{2}}\right)(19+28\alpha^{2})
\nonumber\\ &&\ \ \ \left.
-\frac{1}{288}\left(\frac{V_{0}^{(6)}}{V_{0}^{''}}\right)(5+4\alpha^{2})\right\}.
\end{eqnarray}
Here
\begin{eqnarray}
\alpha&=&n+\frac{1}{2},\ n=\left\{
\begin{array}{l}
0,1,2,\cdots,\ {\rm Re}(E)>0\\ -1,-2,-3,\cdots,\ {\rm Re}(E)<0,
\end{array}
\right.\\
V_{0}^{(n)}&=&\left.\frac{d^{n}V}{dr_{\ast}^{n}}\right|_{r_{\ast}=r_{\ast}(r_{max})}.
\end{eqnarray}

\begin{figure}[!]
\includegraphics{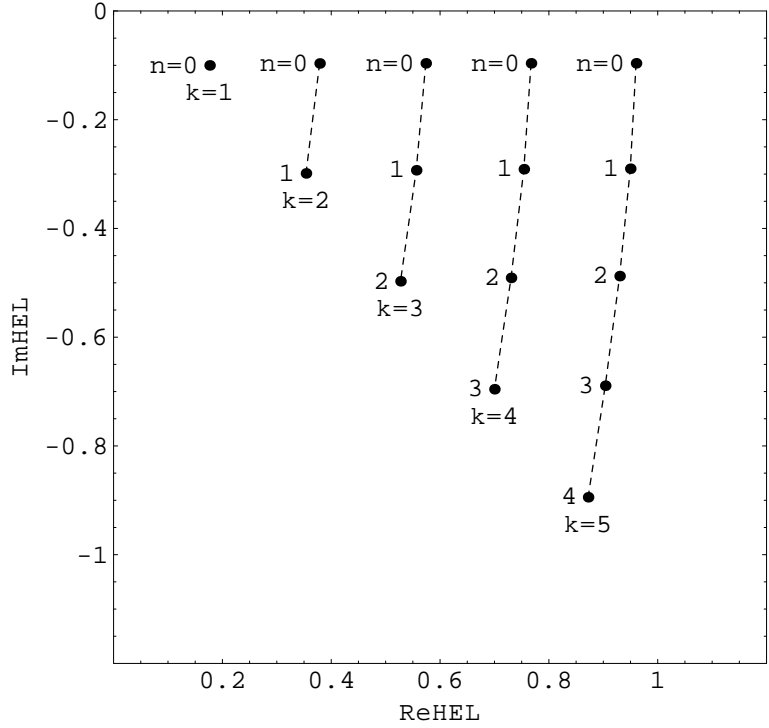}
\caption{\label{masslessqnm} Massless Dirac quasi-normal mode
frequencies}
\end{figure}

Plugging the effective potential in Eq.~(\ref{m0V}) into the
formula above, we obtain the complex quasi-normal mode frequencies
for the massless Dirac field. The values for $0\leq n<\kappa$ are
listed in Table~\ref{table1}. The values for negative $n$ are
related to those with positive $n$ by a reflection of the
imaginary axis so we do not list them out here. The quasi-normal
mode frequencies for positive $\kappa$ are plotted in
Fig.~\ref{masslessqnm}. The variations of the mode frequencies
here are similar to those with integral spin fields in the
Schwarzschild black hole spacetime \cite{Iyer2}. The real part
Re$(E)$ of the frequencies decreases as the mode number $n$
increases for the same angular momentum number $\kappa$. On the
other hand, the magitude of the imaginary part Im$(E)$ of the
frequencies increases with $n$. This indicates that the
quasi-normal modes with higher mode numbers decay faster than the
low-lying ones. Hence, the low-lying quasi-normal modes, possibly
with different values of $|\kappa|$, are most relevant to the
description of the evolution of a massless Dirac field in the
Schwarzschild black hole spacetime.

\section{Quasi-normal mode frequencies for the massive Dirac
field}

In this section we shall evaluate the quasi-normal mode
frequencies of the massive Dirac field. Again we start with the
radial equation (Eq.~(\ref{V1})),
\begin{equation}
\left(-\frac{d^{2}}{d\hat{r}_{\ast}^{2}}+V(r,\kappa,m,E)\right)\hat{F}
=E^{2}\hat{F},
\end{equation}
where $\hat{r}_{\ast}=r+2\ {\rm ln}(r/2-1)+(1/2E){\rm
tan}^{-1}(mr/\kappa)$, and
\begin{eqnarray}
&&V(r,\kappa,m,E)\nonumber\\
&=&\frac{\Delta^{1/2}(\kappa^{2}+m^{2}r^{2})^{3/2}}
{(r^{2}(\kappa^{2}+m^{2}r^{2})+m\kappa\Delta/2E)^{2}}
\left[\Delta^{1/2}(\kappa^{2}+m^{2}r^{2})^{3/2}+((r-1)(\kappa^{2}+m^{2}r^{2})+
3m^{2}r\Delta)\right]\nonumber\\ &&\ \ \
-\frac{\Delta^{3/2}(\kappa^{2}+m^{2}r^{2})^{5/2}}{(r^2(\kappa^{2}+m^{2}r^{2})+m\kappa\Delta/2E)^{3}}
\left[2r(\kappa^{2}+m^{2}r^{2})+2m^{2}r^{3}+m\kappa(r-1)/E\right].
\label{massiveV}
\end{eqnarray}
Note that the effective potential depends not only on $m$ but also
on the energy $E$. This will complicate our evaluation of the
quasi-normal mode frequencies using the WKB approximation formula
in Eq.~(\ref{WKBeq}) because there are $E$ dependence on both
sides of the equation.

\begin{figure}[!]
\includegraphics{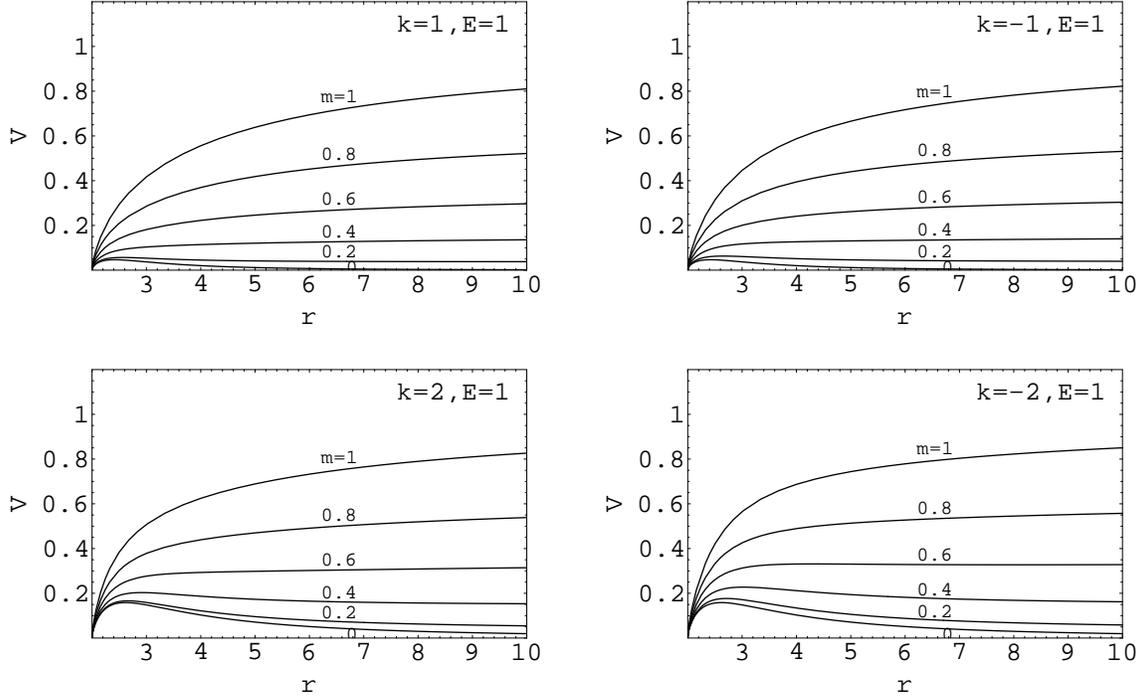}
\caption{\label{vandm} Variation of the effective potential for
the Dirac field with mass $m$ for $\kappa = 1,2,-1$ and $-2$.}
\end{figure}

\subsection{Properties of the effective potential}

Here we analyse the dependence of the effective potential on the
parameters $m$, $\kappa$, and $E$. First, its dependence on $m$ is
showed in Fig.~\ref{vandm} with energy $E=1$ and with
$\kappa=1,2,-1$ and $-2$. For small values of $m$, the potential
is still in the form of a barrier, but with the asymptotic value
\begin{equation}
V(r\rightarrow\infty)=m^2.
\end{equation}
As $m$ is increased, the peak of the potential also increases but
does so very slowly. Eventually, the height of the peak is lower
than the asymptotic value $m^2$. In this case there will be no
quasi-normal modes because for tunneling to occur $E^{2}$ of the
Dirac field must be smaller than the peak value of the potential.
This cannot happen since the energy of the Dirac field is always
larger than its mass $m$. When $m$ is increased further, the peak
disappears all together, and the potential barrier turns
effectively into a potential step. Note that there is a caveat
here. The energy $E$ is unknown at the start of the calculation
(here we have taken a typical value $E=1$) and has to be
determined self-consistently. Hence, Fig.~\ref{vandm} (similarly
for the other figures below) should only be taken as indicative of
the general behaviors of the effective potential $V$.

\begin{figure}[!]
\includegraphics{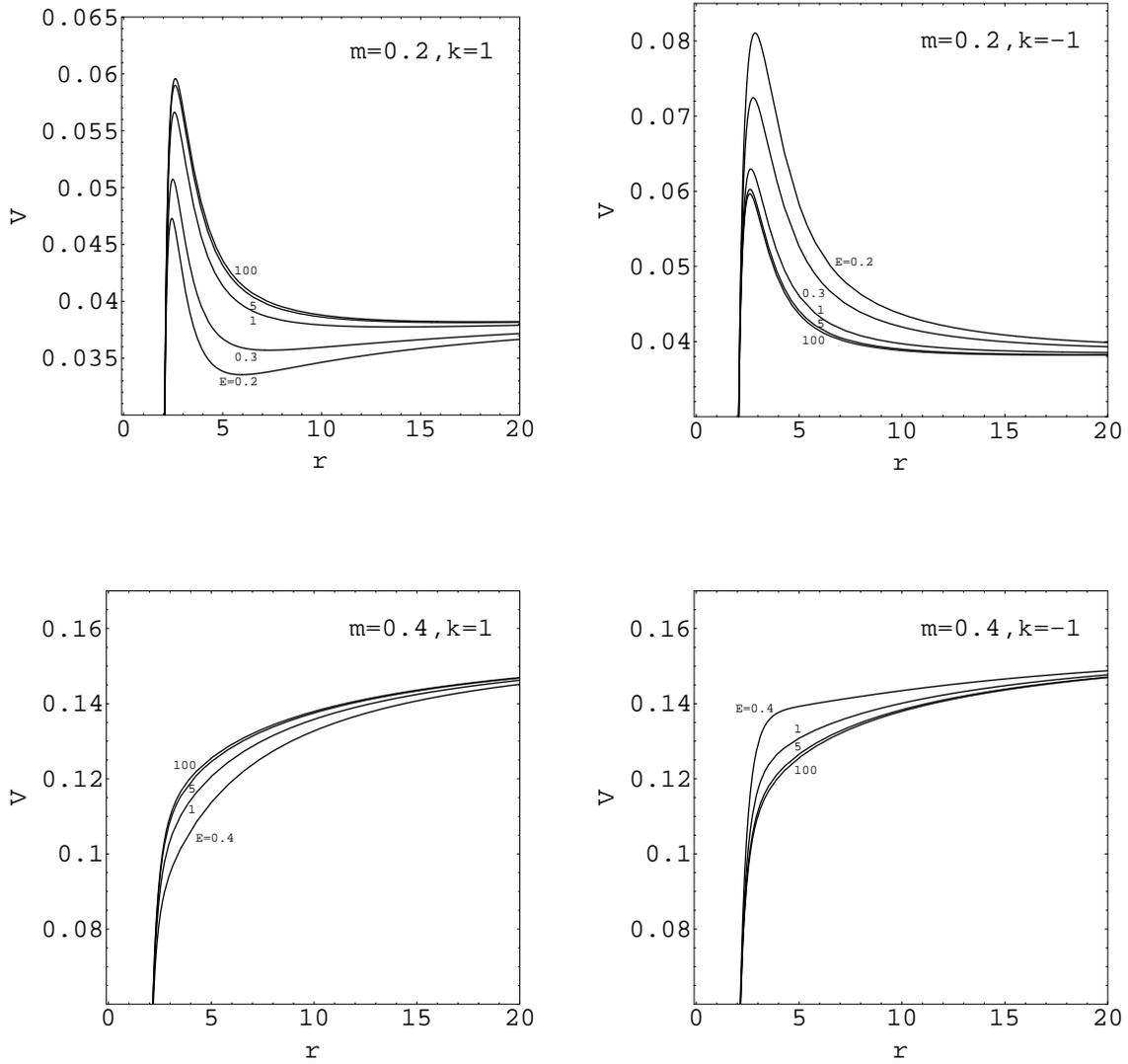}
\caption{\label{vande} Variation of the effective potential for
the Dirac field with energy $E$ for $m = 0.2$ and $0.4$, and
$\kappa = 1$ and $-1$.}
\end{figure}

The effective potential for the massive Dirac field also depends
on the energy $E$, and we show its dependence in Fig.~\ref{vande}.
$E$ is varied from its minimum value $m$. For positive $\kappa$,
the value of the potential increases with $E$, while for negative
$\kappa$, it decreases. The variations become smaller and smaller,
and we can see from the figure that the potentials are nearly
indistingusable for $E=5$ and $E=100$ or higher. The general
behaviors of the potential remain the same for all these values of
$E$. Indeed, from the form of the potential in
Eq.~(\ref{massiveV}), $E$ appears all in the denominators. The
terms involving $E$ can never get large enough to change the
general behaviors of the potential since $E$ cannot be smaller
than $m$.

\begin{table}
\caption{\label{table2}Estimation of the maximum values of the
mass $m$ and $\mu$ (= $m/\kappa$) of the Dirac field above which
quasi-normal modes cannot occur.}
\begin{ruledtabular}
\begin{tabular}{ccccc}
$|\kappa|$&$m_{max}$(positive $\kappa$)&$\mu_{max}$(positive
$\kappa$)&$m_{max}$(negative $\kappa$)&$\mu_{max}$(negative
$\kappa$)\\ \hline 1 & 0.224 & 0.224 & 0.333 & -0.333 \\ 2 & 0.453
& 0.226 & 0.572 & -0.286 \\ 3 & 0.696 & 0.232 & 0.819 & -0.273 \\
4 & 0.944 & 0.236 & 1.067 & -0.267 \\ 5 & 1.192 & 0.238 & 1.316 &
-0.263 \\
\end{tabular}
\end{ruledtabular}
\end{table}

From the dependences of the potential with $m$ and $E$, we can
also estimate the maximum values of $m$ above which quasi-normal
modes cannot occur. From the discussions above, we see that
quasi-normal modes exist only when (i) the peak value of the
potential $V(r=r_{\rm max})$ is larger than $m^2$, and (ii)
$E^{2}$ of the field is smaller than this peak value. Therefore,
one can estimate this maximum value $m_{max}$ from the relation
\begin{equation}
V(r_{max},m_{max},\kappa,E=m_{max})=(m_{max})^{2},
\end{equation}
which can be solved numerically. The result is tabulated in
Table~\ref{table2}, where we have also given the maximum values of
$\mu=m/\kappa$ which will be useful in the calculation below.

\begin{figure}[!]
\includegraphics{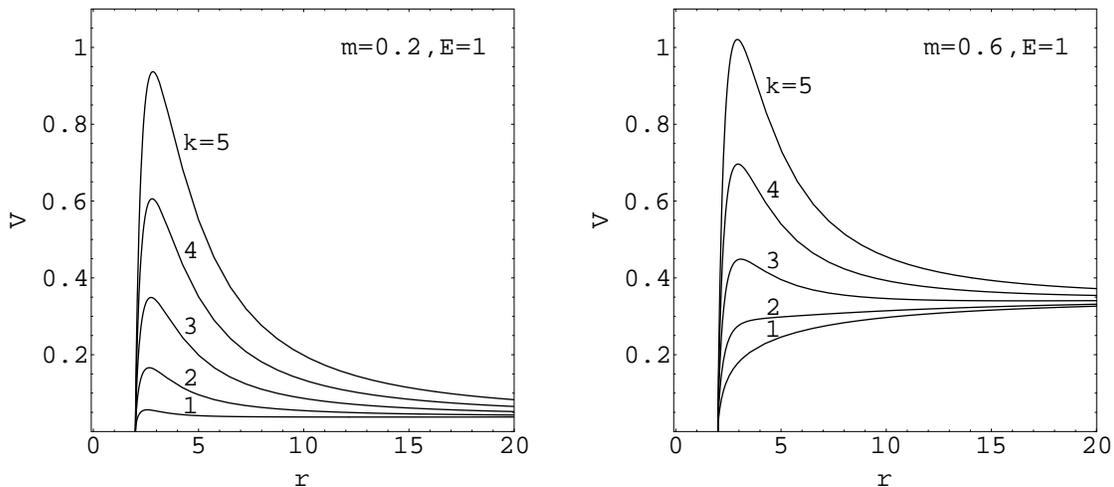}
\caption{\label{vandk} Variation of the effective potential for
the Dirac field with $\kappa$ for $m=0.2$ and $0.6$.}
\end{figure}

Finally, in Fig.~\ref{vandk} we show the dependence of the
effective potential with the angular momentum quantum number
$\kappa$. We see that as $\kappa$ increases, the behaviors of the
potential approach to that of the massless one, as shown in
Fig.~\ref{m0kappa1}, regardless the mass of the field. This can be
understood by writing the potential in Eq.~(\ref{massiveV}) as
\begin{eqnarray}
&&V(r,\kappa,m=\kappa\mu,E)\nonumber\\
&=&\frac{|\kappa|\Delta^{1/2}(1+\mu^{2}r^{2})^{3/2}}
{(r^{2}(1+\mu^{2}r^{2})+\mu\Delta/2E)^{2}}
\left[|\kappa|\Delta^{1/2}(1+\mu^{2}r^{2})^{3/2}+((r-1)(1+\mu^{2}r^{2})+
3\mu^{2}r\Delta)\right]\nonumber\\ &&\ \ \
-\frac{|\kappa|\Delta^{3/2}(1+\mu^{2}r^{2})^{5/2}}{(r^2(1+\mu^{2}r^{2})+
\mu\Delta/2E)^{3}}
\left[2r(1+\mu^{2}r^{2})+2\mu^{2}r^{3}+\mu(r-1)/E\right],
\label{alphaV}
\end{eqnarray}
where $\mu$ is as defined above. The dependence of mass is all
through $\mu$. As $|\kappa|\rightarrow\infty$, $\mu\rightarrow0$
regardless the value of $m$, and the effective potential will
approach to that of the massless one. Also from
Table~\ref{table2}, the maximum magnitudes of $\mu$ are around
$0.2\sim 0.3$ regardless the value of $\kappa$. $\mu$ can thus be
used as a small parameter for expansions as we shall do below.

\subsection{Evaluation of the quasi-normal mode frequencies}

The evaluation of the massive Dirac quasi-normal mode frequencies
is more complicated than the massless ones due to the presence of
the energy dependence on both sides of the Eq.~(\ref{WKBeq}).
Since the maximum magnitudes of $\mu$ are only around $0.2\sim
0.3$, we can try to obtain the quasi-normal mode frequencies in
the WKB approximation as power series of $\mu$ for given values of
$\kappa$ similar to the power series expansions in \cite{Seidel}
and \cite{Simone}.

We first express the position of the peak of the effective
potential as a series up to order $\mu^{6}$,
\begin{eqnarray}
r_{max}&=&r_{0}+r_{1}\mu+r_{2}\mu^{2}+r_{3}\mu^{3}+r_{4}\mu^{4}+r_{5}\mu^{5}+r_{6}
\mu^{6}\nonumber\\ &=& r_{0}+\Sigma,
\end{eqnarray}
by requiring
\begin{eqnarray}
0&=& V^{'}(r_{max})\nonumber\\ &=& V^{'}(r_{0})+\Sigma
V^{''}(r_{0})+\frac{1}{2}\Sigma^{2}V^{'''}(r_{0})+\frac{1}{6}\Sigma^{3}V^{(4)}(r_{0})\nonumber\\
&&+\frac{1}{24}\Sigma^{4}V^{(5)}(r_{0})+\frac{1}{120}\Sigma^{5}V^{(6)}(r_{0})
+\frac{1}{720}\Sigma^{6}V^{(7)}(r_{0}).
\end{eqnarray}
Since $r_{0}$ is just the position of the peak for the massless
case obtained in Section~\ref{section3}, we can evaluate the
coefficients $r_{i}$'s order by order from this equation. For
example, for $|\kappa|=1$, we obtain
\begin{eqnarray}
r_{max}&=&2.42-\mu\left(\frac{0.182}{E}\right)+\mu^{2}\left(3.66+\frac{0.0755}{E^{2}}\right)
-\mu^{3}\left(\frac{1.07}{E}+\frac{0.0322}{E^{3}}\right)\nonumber\\
&&+\mu^{4}\left(18.8-\frac{0.252}{E^{2}}
+\frac{0.0142}{E^{4}}\right)-\mu^{5}\left(\frac{0.526}{E}-\frac{0.624}{E^{3}}
+\frac{0.00638}{E^{5}}\right)\nonumber\\ &&\ \ \
+\mu^{6}\left(117-\frac{8.16}{E^{2}}-\frac{0.597}{E^{4}}
+\frac{0.00293}{E^{6}}\right).\label{rmaxexpand}
\end{eqnarray}
The coefficients in the above expansion involve the unknown $E$.
We can likewise expand $E$,
\begin{equation}
E=E_{0}+E_{1}\mu+E_{2}\mu^{2}+E_{3}\mu^{3}+E_{4}\mu^{4}+E_{5}\mu^{5}+E_{6}\mu^{6}.
\end{equation}
We plug this expansion back into Eq.~(\ref{rmaxexpand}), and
expand again up to order $\mu^{6}$. Using this new expansion for
$r_{max}$, we can then expand the derivative of the potential
$V^{(n)}_{0}$. Note that the derivatives are performed with
respect to $\hat{r}_{\ast}$, and
\begin{eqnarray}
\frac{d}{d\hat{r}_{\ast}}&=&\left[\frac{1}{1-2/r}+\frac{\mu}
{2E(1+\mu^{2}r^{2})}\right]^{-1}\frac{d}{dr}\nonumber\\ &=&
\left[\left(1-\frac{2}{r}\right)-\frac{1}{2E}\left(1-\frac{2}{r}\right)^{2}\mu
+\frac{1}{4E^{2}}\left(1-\frac{2}{r}\right)^{3}\mu^{2}\right.\nonumber\\
&&\ \
-\frac{1}{8E^{3}r^{2}}\left(1-\frac{2}{r}\right)^{2}(4-4r+r^{2}-4E^{2}r^{4})\mu^{3}\nonumber\\
&&\ \
+\frac{1}{16E^{4}r^{2}}\left(1-\frac{2}{r}\right)^{3}(4-4r+r^{2}-8E^{2}r^{4})\mu^{4}\nonumber\\
&&\ \
-\frac{1}{32E^{5}r^{4}}\left(1-\frac{2}{r}\right)^{3}(16-32r+24r^{2}-8r^{3}\nonumber\\
&&\ \ \ \ \ +(1-48E^{2})r^{4}
+48E^{2}r^{5}-12E^{2}r^{6}+16E^{4}r^{8})\mu^{5}\nonumber\\ &&\ \
+\frac{1}{64E^{6}r^{4}}\left(1-\frac{2}{r}\right)^{3}(16-32r+24r^{2}-8r^{3}\nonumber\\
&&\ \ \ \ \ \ +(1-64E^{2})r^{4}
+64E^{2}r^{5}-16E^{2}r^{6}+48E^{4}r^{8})\mu^{6}+\cdots\Bigg]\frac{d}{dr}.
\end{eqnarray}
We put all these expansions back to Eq.~(\ref{WKBeq}) and solve
the coefficients $E_{i}$'s self-consistently order by order in
$\mu$. The results are tabulated in Tables~\ref{table3} and
\ref{table4}.

\begin{table}
\caption{\label{table3}Real parts of the coefficients of the
expansions in powers of $\mu$ for the quasi-normal mode
frequencies with $|\kappa|=1$ to $5$.}
\begin{ruledtabular}
\begin{tabular}{ccccccccc}
$|\kappa|$ & $n$ & Re($E_{0}$) & Re($E_{1}$) & Re($ E_{2}$) & Re($
E_{3}$) & Re($ E_{4}$) & Re($ E_{5}$) & Re($E_{6}$)\\ \hline 1 & 0
& 0.176 & -0.168 & 0.521 & -0.393 & -0.00367 & -3.13 & 74.2 \\ 2 &
0 & 0.379 & -0.166 & 1.49 & 0.158 & 1.01 & -0.728 & -0.845\\ & 1 &
0.354 & -0.136 & 0.478 & -0.510 & -3.81 & 9.23 & -24.0
\\ 3 & 0 & 0.574 & -0.166 & 2.43 & 0.335 & 1.84 & 0.721 & 2.75
\\ & 1 & 0.556 & -0.152 & 1.55 & -0.366 & -1.92 & -3.30 &
-8.91
\\ & 2 & 0.527 & -0.129 & 0.463 & -0.455 & -5.84 & 16.6 &
-93.0
\\ 4 & 0 & 0.767 & -0.166 & 3.34 & 0.404 & 2.56 & 1.13 & 6.85
\\ & 1 & 0.754 & -0.158 & 2.61 & -0.115 & 0.291 & -2.31 & -2.43
\\ & 2 & 0.730 & -0.144 & 1.50 & -0.494 & -5.64 & -0.255 &
-49.8
\\ & 3 & 0.700 & -0.126 & 0.460 & -0.427 & -7.39 & 20.8 &
-152
\\ 5 & 0 & 0.960 & -0.166 & 4.23 & 0.437 & 3.24 & 1.28 & 9.98
\\ & 1 & 0.950 & -0.161 & 3.62 & 0.0610 & 1.80 & -0.952 & 2.32\\
& 2 & 0.930 & -0.152 & 2.60 & -0.358 & -3.02 & -3.28 & -23.4
\\ & 3 & 0.904 & -0.138 & 1.46 & -0.514 & -8.82 & 4.03 & -103\\
& 4 & 0.872 & -0.124 & 0.462 & -0.408 & -8.81 & 23.5 & -204
\\
\end{tabular}
\end{ruledtabular}
\end{table}

\begin{table}
\caption{\label{table4}Imaginary parts of the coefficients of the
expansions in powers of $\mu$ for the quasi-normal mode
frequencies with $|\kappa|=1$ to $5$.}
\begin{ruledtabular}
\begin{tabular}{ccccccccc}
$|\kappa|$ & $n$ & Im($E_{0}$) & Im($E_{1}$) & Im($ E_{2}$) & Im($
E_{3}$) & Im($ E_{4}$) & Im($ E_{5}$) & Im($E_{6}$)\\ \hline 1 & 0
& -0.100 & -0.0711 & 0.452 & 0.375 & 1.38 & -8.59 & -14.0 \\ 2 & 0
& -0.0965 & -0.0282 & 0.653 & 0.480 & 1.38 & 1.59 & -3.47
\\ & 1 & -0.299 & -0.0771 & 1.16 & -0.0876 & -5.43 & -16.9 &
7.54
\\ 3 & 0 & -0.0963 & -0.0185 & 0.690 & 0.368 & 1.21 & 1.46 &
4.76
\\ & 1 & -0.293 & -0.0533 & 1.64 & 0.391 & 1.38 & -4.30 &
22.6
\\ & 2 & -0.497 & -0.0770 & 1.72 & -0.172 & -13.3 & -18.9 & 37.7
\\ 4 & 0 & -0.0963 & -0.0139 & 0.704 & 0.291 & 1.13 & 1.14 &
6.02 \\ & 1 & -0.291 & -0.0406 & 1.85 & 0.516 & 2.87 & 0.442 &
21.0 \\ & 2 & -0.491 & -0.0627 & 2.38 & 0.186 & -2.21 & -11.0 &
59.4 \\ & 3 & -0.696 & -0.0771 & 2.27 & -0.204 & -20.5 & -19.6 &
52.6 \\ 5 & 0 & -0.0963 & -0.0111 & 0.710 & 0.239 & 1.08 & 0.914 &
6.23 \\ & 1 & -0.290 & -0.0327 & 1.96 & 0.517 & 3.20 & 1.60 & 19.9
\\ & 2 & -0.488 & -0.0521 & 2.76 & 0.399 & 2.25 & -4.00 &
53.6 \\ & 3 & -0.689 & -0.0672 & 3.01 & 0.0573 & -7.44 & -15.0 &
81.5 \\ & 4 & -0.894 & -0.0773 & 2.80 & -0.223 & -27.5 & -19.8 &
53.6 \\
\end{tabular}
\end{ruledtabular}
\end{table}

\begin{figure}[!]
\includegraphics{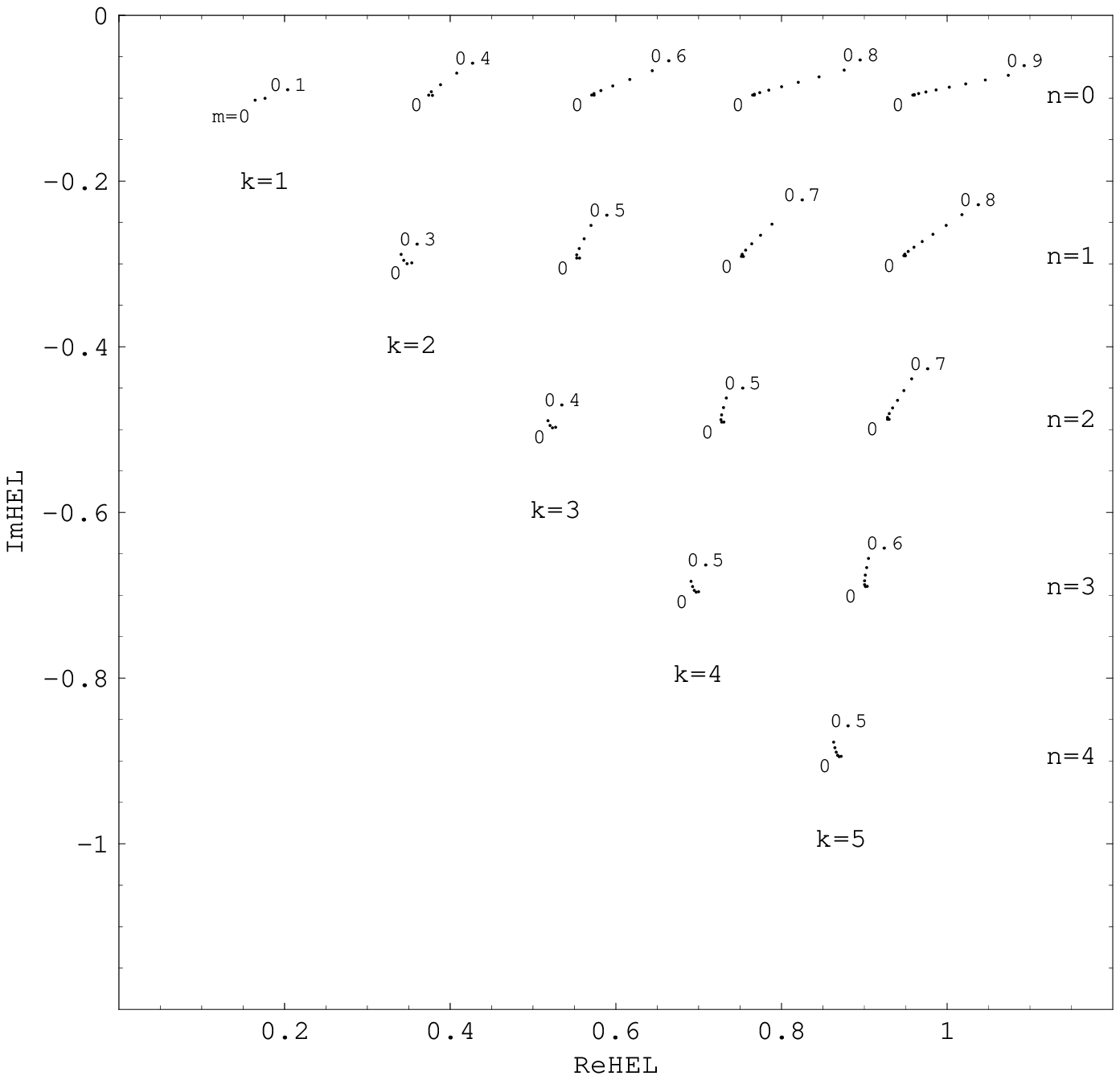}
\caption{\label{finalplot} Massive Dirac quasi-normal mode
frequencies}
\end{figure}

With these coefficients we can evaluate the quasi-normal mode
frequencies by putting in the value of $\mu$ for both positive and
negative values of $\kappa$. From Table~\ref{table2}, the maximum
value of $\mu$ that could have quasi-normal modes is 0.224 for
$\kappa=1$ and increases to 0.238 for $\kappa=5$. In fact, the
value will increase for larger values of $\kappa$, but converges
to the value $\sim 0.25$. Similarly for the variation of
$\mu_{max}$ with negative $\kappa$, the magnitude decreases from
0.333 for $\kappa=-1$ to 0.263 for $\kappa=-5$, and will converge
to $\sim 0.25$ for large values of $\kappa$. Therefore, we shall
try to evaluate the frequencies for $|\mu|$ from 0 to 0.2.

First we examine the frequencies for $\kappa=1$ and $n=0$. For
$m=0$, we have
\begin{equation}
E=E_{0}=0.176-0.100i,
\end{equation}
which is the massless case we have obtained in the last section.

For $m=0.1$, we have
\begin{eqnarray}
E_{0}&=&0.176-0.100i,\nonumber\\
E_{1}\mu&=&-0.0168-0.00711i,\nonumber\\
E_{2}\mu^{2}&=&0.00521+0.00452i,\nonumber\\
E_{3}\mu^{3}&=&-0.000393+0.000375i,\nonumber\\
E_{4}\mu^{4}&=&-0.00000367+0.000138i\nonumber\\
E_{5}\mu^{5}&=&-0.0000313-0.0000859i,\nonumber\\
E_{6}\mu^{6}&=&0.0000742-0.0000140i,
\end{eqnarray}
and
\begin{equation}
E=0.164-0.102i\label{em1n0}
\end{equation}
Hence, if we take three significant figures, both the real and the
imaginary parts of $E_{3}\mu^{3}$ and higher order terms do not
contribute to the frequency. The result in Eq.~(\ref{em1n0}) is
thus accurate up to at least three significant figures.

For $m=0.2$, we have
\begin{eqnarray}
E_{0}&=&0.176-0.100i,\nonumber\\
E_{1}\mu&=&-0.0337-0.0142i,\nonumber\\
E_{2}\mu^{2}&=&0.0208+0.0181i,\nonumber\\
E_{3}\mu^{3}&=&-0.00315+0.00300i,\nonumber\\
E_{4}\mu^{4}&=&-0.00000587+0.00221i\nonumber\\
E_{5}\mu^{5}&=&-0.00100-0.00275i,\nonumber\\
E_{6}\mu^{6}&=&0.00475-0.000893i,
\end{eqnarray}
and
\begin{equation}
E=0.164-0.0947i\label{em2n0}
\end{equation}
Here we see that even the highest order term $E_{6}\mu^{6}$ will
contribute to both the real and the imaginary parts of $E$. If we
require our result to be accurate up to three significant figures,
we need to go to orders higher than $\mu^{6}$. However, we have
only expanded terms up to $\mu^{6}$, we are thus forced to
disregard this result for $m=0.2$, even though there should be a
quasi-normal mode around that value.

In summary, for different values of $\kappa$ and $n$, we increase
the values of $m$ (or $\mu$) from 0 up to the point that three
significant figure accuracy cannot be maintained with terms up to
only $\mu^{6}$, and then we stop at that value. The results for
$\kappa=1$ to $5$ are plotted in Fig.~\ref{finalplot}. Comparing
with Table \ref{table2}, we see that the largest value of $m$ for
each $\kappa$ and $n$ that we present in Fig.~\ref{finalplot} is
in fact very close to the maximum value $m_{max}$. On the other
hand, the trends of the variations of $E$ with $m$ are similar for
negative $\kappa$ so we do not plot them out.

As we can see from Fig.~\ref{finalplot}, the real parts of the
quasi-normal mode frequencies in general increase with the mass of
the field. This can be understood from the fact that when the mass
of the field increases, the peak of the effective potential gets
higher. The real part of the energy should therefore increase. We
also see that the magnitudes of the imaginary parts decrease with
the mass. We thus expect Dirac fields with larger masses (same
$\kappa$ and $n$) to decay slower during their evolutions.

\section{Conclusions and Discussions}

We have evaluated both the massless and the massive Dirac
quasi-normal mode frequencies in the Schwarzschild black hole
spacetime using the WKB approximation. Only the cases with
$\kappa>n$, where $\kappa$ is the total angular momentum number
and n is the mode number, are considered since these are the cases
known to be accurate for the WKB approximation with the integral
spin fields \cite{Iyer2}. For the massless Dirac field, we see
that the real parts of the quasi-normal mode frequencies increase
with $\kappa$, but they decrease with $n$ for fixed $\kappa$. As
for the imaginary parts, the magnitudes of which increase with
$n$. Hence, the fields with $n=0$ will decay most slowly.

For the massive cases, the evaluation of the quasi-normal mode
frequencies is complicated by the fact that the effective
potential also depends on the energy $E$. A further approximation
is needed to carry through the calculation in which we make
perturbative expansions for all the relevant quantities in powers
of the parameter $\mu=m/\kappa$ where $m$ is the mass of the
field. In this way we are able to obtain the numerical values of
both the real and the imaginary parts of the mode frequencies up
to three significant figures. In general, when the mass of the
field is increased, the real parts of the frequencies increase
while the magnitudes of the imaginary parts decrease. Fields with
higher masses will therefore decay slower. This result is similar
to that of the massive scalar field \cite{Simone}.

The calculation carried out here can be extended to other black
hole spacetimes, for example, the charged Reissner-Nordstr\"om
black hole or the rotating Kerr black hole. However, for the Kerr
black hole, the situation is further complicated by the fact that
there is no close form for the effective potential. Furthermore,
we can consider dilaton black holes or black holes in the anti-de
Sitter space both of which are closely related to string theories.

As we have mentioned ealier, our main concern is about the
evolution of Dirac fields in curved spacetimes, especially black
hole spacetimes. After calculating the quasi-normal mode
frequencies, our next step would be to understand the late-time
tail behaviors of the fields. Together with the Green's function
method, we should be able to get a more complete picture of the
evolution, and hopefully towards a better understanding of the
gravitational collapse of Dirac (or fermionic) matter.

\begin{acknowledgments}
This work is supported by the National Science Council of the
Republic of China under contract number NSC 90-2112-M-032-004.
\end{acknowledgments}

\bibliography{DiracQNM}

\begin{thebibliography}{22}
\expandafter\ifx\csname natexlab\endcsname\relax\def\natexlab#1{#1}\fi
\expandafter\ifx\csname bibnamefont\endcsname\relax
  \def\bibnamefont#1{#1}\fi
\expandafter\ifx\csname bibfnamefont\endcsname\relax
  \def\bibfnamefont#1{#1}\fi
\expandafter\ifx\csname citenamefont\endcsname\relax
  \def\citenamefont#1{#1}\fi
\expandafter\ifx\csname url\endcsname\relax
  \def\url#1{\texttt{#1}}\fi
\expandafter\ifx\csname urlprefix\endcsname\relax\def\urlprefix{URL }\fi
\providecommand{\bibinfo}[2]{#2}
\providecommand{\eprint}[2][]{\url{#2}}

\bibitem[{\citenamefont{Finster
  et~al.}(1999{\natexlab{a}})\citenamefont{Finster, Smoller, and
  Yau}}]{Finster1}
\bibinfo{author}{\bibfnamefont{F.}~\bibnamefont{Finster}},
  \bibinfo{author}{\bibfnamefont{J.}~\bibnamefont{Smoller}}, \bibnamefont{and}
  \bibinfo{author}{\bibfnamefont{S.-T.} \bibnamefont{Yau}},
  \bibinfo{journal}{Phys. Rev. D.} \textbf{\bibinfo{volume}{59}},
  \bibinfo{pages}{104020} (\bibinfo{year}{1999}{\natexlab{a}}).

\bibitem[{\citenamefont{Finster
  et~al.}(1999{\natexlab{b}})\citenamefont{Finster, Smoller, and
  Yau}}]{Finster2}
\bibinfo{author}{\bibfnamefont{F.}~\bibnamefont{Finster}},
  \bibinfo{author}{\bibfnamefont{J.}~\bibnamefont{Smoller}}, \bibnamefont{and}
  \bibinfo{author}{\bibfnamefont{S.-T.} \bibnamefont{Yau}},
  \bibinfo{journal}{Phys. Lett.} \textbf{\bibinfo{volume}{A259}},
  \bibinfo{pages}{431} (\bibinfo{year}{1999}{\natexlab{b}}).

\bibitem[{\citenamefont{Finster
  et~al.}(1999{\natexlab{c}})\citenamefont{Finster, Smoller, and
  Yau}}]{Finster3}
\bibinfo{author}{\bibfnamefont{F.}~\bibnamefont{Finster}},
  \bibinfo{author}{\bibfnamefont{J.}~\bibnamefont{Smoller}}, \bibnamefont{and}
  \bibinfo{author}{\bibfnamefont{S.-T.} \bibnamefont{Yau}},
  \bibinfo{journal}{Commun. Math. Phys.} \textbf{\bibinfo{volume}{205}},
  \bibinfo{pages}{249} (\bibinfo{year}{1999}{\natexlab{c}}).

\bibitem[{\citenamefont{Finster
  et~al.}(2000{\natexlab{a}})\citenamefont{Finster, Smoller, and
  Yau}}]{Finster4}
\bibinfo{author}{\bibfnamefont{F.}~\bibnamefont{Finster}},
  \bibinfo{author}{\bibfnamefont{J.}~\bibnamefont{Smoller}}, \bibnamefont{and}
  \bibinfo{author}{\bibfnamefont{S.-T.} \bibnamefont{Yau}},
  \bibinfo{journal}{J. Math. Phys.} \textbf{\bibinfo{volume}{41}},
  \bibinfo{pages}{2173} (\bibinfo{year}{2000}{\natexlab{a}}).

\bibitem[{\citenamefont{Finster
  et~al.}(2000{\natexlab{b}})\citenamefont{Finster, Kamran, Smoller, and
  Yau}}]{Finster5}
\bibinfo{author}{\bibfnamefont{F.}~\bibnamefont{Finster}},
  \bibinfo{author}{\bibfnamefont{N.}~\bibnamefont{Kamran}},
  \bibinfo{author}{\bibfnamefont{J.}~\bibnamefont{Smoller}}, \bibnamefont{and}
  \bibinfo{author}{\bibfnamefont{S.-T.} \bibnamefont{Yau}},
  \bibinfo{journal}{Commun. Pure Appl. Math.} \textbf{\bibinfo{volume}{53}},
  \bibinfo{pages}{902} (\bibinfo{year}{2000}{\natexlab{b}}).

\bibitem[{\citenamefont{Finster
  et~al.}(2000{\natexlab{c}})\citenamefont{Finster, Smoller, and
  Yau}}]{Finster6}
\bibinfo{author}{\bibfnamefont{F.}~\bibnamefont{Finster}},
  \bibinfo{author}{\bibfnamefont{J.}~\bibnamefont{Smoller}}, \bibnamefont{and}
  \bibinfo{author}{\bibfnamefont{S.-T.} \bibnamefont{Yau}},
  \bibinfo{journal}{J. Math. Phys.} \textbf{\bibinfo{volume}{41}},
  \bibinfo{pages}{3943} (\bibinfo{year}{2000}{\natexlab{c}}).

\bibitem[{\citenamefont{Finster
  et~al.}(2002{\natexlab{a}})\citenamefont{Finster, Smoller, and
  Yau}}]{Finster7}
\bibinfo{author}{\bibfnamefont{F.}~\bibnamefont{Finster}},
  \bibinfo{author}{\bibfnamefont{J.}~\bibnamefont{Smoller}}, \bibnamefont{and}
  \bibinfo{author}{\bibfnamefont{S.-T.} \bibnamefont{Yau}},
  \bibinfo{journal}{Adv. Theor. Math. Phys.} \textbf{\bibinfo{volume}{4}},
  \bibinfo{pages}{1231} (\bibinfo{year}{2002}{\natexlab{a}}).

\bibitem[{\citenamefont{Finster et~al.}({\natexlab{a}})\citenamefont{Finster,
  Smoller, and Yau}}]{Finster8}
\bibinfo{author}{\bibfnamefont{F.}~\bibnamefont{Finster}},
  \bibinfo{author}{\bibfnamefont{J.}~\bibnamefont{Smoller}}, \bibnamefont{and}
  \bibinfo{author}{\bibfnamefont{S.-T.} \bibnamefont{Yau}},
  \eprint{gr-qc/0211043}.

\bibitem[{\citenamefont{Finster
  et~al.}(2002{\natexlab{b}})\citenamefont{Finster, Kamran, Smoller, and
  Yau}}]{Finster9}
\bibinfo{author}{\bibfnamefont{F.}~\bibnamefont{Finster}},
  \bibinfo{author}{\bibfnamefont{N.}~\bibnamefont{Kamran}},
  \bibinfo{author}{\bibfnamefont{J.}~\bibnamefont{Smoller}}, \bibnamefont{and}
  \bibinfo{author}{\bibfnamefont{S.-T.} \bibnamefont{Yau}},
  \bibinfo{journal}{Commun. Math. Phys.} \textbf{\bibinfo{volume}{230}},
  \bibinfo{pages}{201} (\bibinfo{year}{2002}{\natexlab{b}}).

\bibitem[{\citenamefont{Finster et~al.}({\natexlab{b}})\citenamefont{Finster,
  Kamran, Smoller, and Yau}}]{Finster10}
\bibinfo{author}{\bibfnamefont{F.}~\bibnamefont{Finster}},
  \bibinfo{author}{\bibfnamefont{N.}~\bibnamefont{Kamran}},
  \bibinfo{author}{\bibfnamefont{J.}~\bibnamefont{Smoller}}, \bibnamefont{and}
  \bibinfo{author}{\bibfnamefont{S.-T.} \bibnamefont{Yau}},
  \eprint{gr-qc/0005088}.

\bibitem[{\citenamefont{Frolov and Novikov}(1998)}]{Frolov}
\bibinfo{author}{\bibfnamefont{V.~P.} \bibnamefont{Frolov}} \bibnamefont{and}
  \bibinfo{author}{\bibfnamefont{I.~D.} \bibnamefont{Novikov}},
  \emph{\bibinfo{title}{Black Hole Physics: Basic Concepts and New
  Developments}} (\bibinfo{publisher}{Kluwer Academic Publishers},
  \bibinfo{year}{1998}).

\bibitem[{\citenamefont{Kokkotas and Schmidt}(1999)}]{Kokkotas}
\bibinfo{author}{\bibfnamefont{K.~D.} \bibnamefont{Kokkotas}} \bibnamefont{and}
  \bibinfo{author}{\bibfnamefont{B.~G.} \bibnamefont{Schmidt}},
  \bibinfo{journal}{Living Rev. Relativity} \textbf{\bibinfo{volume}{2}}
  (\bibinfo{year}{1999}).

\bibitem[{\citenamefont{Schutz and Will}(1985)}]{Schutz}
\bibinfo{author}{\bibfnamefont{B.~F.} \bibnamefont{Schutz}} \bibnamefont{and}
  \bibinfo{author}{\bibfnamefont{C.~M.} \bibnamefont{Will}},
  \bibinfo{journal}{Astrophys. J. Lett.} \textbf{\bibinfo{volume}{291}},
  \bibinfo{pages}{L33} (\bibinfo{year}{1985}).

\bibitem[{\citenamefont{Iyer and Will}(1987)}]{Iyer1}
\bibinfo{author}{\bibfnamefont{S.}~\bibnamefont{Iyer}} \bibnamefont{and}
  \bibinfo{author}{\bibfnamefont{C.~M.} \bibnamefont{Will}},
  \bibinfo{journal}{Phys. Rev. D.} \textbf{\bibinfo{volume}{35}},
  \bibinfo{pages}{3621} (\bibinfo{year}{1987}).

\bibitem[{\citenamefont{Iyer}(1987)}]{Iyer2}
\bibinfo{author}{\bibfnamefont{S.}~\bibnamefont{Iyer}}, \bibinfo{journal}{Phys.
  Rev. D.} \textbf{\bibinfo{volume}{35}}, \bibinfo{pages}{3632}
  (\bibinfo{year}{1987}).

\bibitem[{\citenamefont{Brill and Wheeler}(1957)}]{Brill}
\bibinfo{author}{\bibfnamefont{D.~R.} \bibnamefont{Brill}} \bibnamefont{and}
  \bibinfo{author}{\bibfnamefont{J.~A.} \bibnamefont{Wheeler}},
  \bibinfo{journal}{Rev. Mod. Phys.} \textbf{\bibinfo{volume}{29}},
  \bibinfo{pages}{465} (\bibinfo{year}{1957}).

\bibitem[{\citenamefont{Bjorken and Drell}(1964)}]{Bjorken}
\bibinfo{author}{\bibfnamefont{J.~D.} \bibnamefont{Bjorken}} \bibnamefont{and}
  \bibinfo{author}{\bibfnamefont{S.~D.} \bibnamefont{Drell}},
  \emph{\bibinfo{title}{Relativistic Quantum Mechanics}}
  (\bibinfo{publisher}{McGraw Hill}, \bibinfo{year}{1964}).

\bibitem[{\citenamefont{Chandrasekhar}(1983)}]{Chandrasekhar}
\bibinfo{author}{\bibfnamefont{S.}~\bibnamefont{Chandrasekhar}},
  \emph{\bibinfo{title}{The Mathematical Theory of Black Holes}}
  (\bibinfo{publisher}{Clarendon Press}, \bibinfo{year}{1983}).

\bibitem[{\citenamefont{Cooper et~al.}(1995)\citenamefont{Cooper, Khare, and
  Sukhatme}}]{Cooper}
\bibinfo{author}{\bibfnamefont{F.}~\bibnamefont{Cooper}},
  \bibinfo{author}{\bibfnamefont{A.}~\bibnamefont{Khare}}, \bibnamefont{and}
  \bibinfo{author}{\bibfnamefont{U.}~\bibnamefont{Sukhatme}},
  \bibinfo{journal}{Phys. Rept.} \textbf{\bibinfo{volume}{251}},
  \bibinfo{pages}{267} (\bibinfo{year}{1995}).

\bibitem[{\citenamefont{Anderson and Price}(1991)}]{Anderson}
\bibinfo{author}{\bibfnamefont{A.}~\bibnamefont{Anderson}} \bibnamefont{and}
  \bibinfo{author}{\bibfnamefont{R.~H.} \bibnamefont{Price}},
  \bibinfo{journal}{Phys. Rev. D.} \textbf{\bibinfo{volume}{43}},
  \bibinfo{pages}{3147} (\bibinfo{year}{1991}).

\bibitem[{\citenamefont{Seidel and Iyer}(1990)}]{Seidel}
\bibinfo{author}{\bibfnamefont{E.}~\bibnamefont{Seidel}} \bibnamefont{and}
  \bibinfo{author}{\bibfnamefont{S.}~\bibnamefont{Iyer}},
  \bibinfo{journal}{Phys. Rev. D.} \textbf{\bibinfo{volume}{41}},
  \bibinfo{pages}{374} (\bibinfo{year}{1990}).

\bibitem[{\citenamefont{Simone and Will}(1992)}]{Simone}
\bibinfo{author}{\bibfnamefont{L.~E.} \bibnamefont{Simone}} \bibnamefont{and}
  \bibinfo{author}{\bibfnamefont{C.~M.} \bibnamefont{Will}},
  \bibinfo{journal}{Class. Quantum Grav.} \textbf{\bibinfo{volume}{9}},
  \bibinfo{pages}{963} (\bibinfo{year}{1992}).

\end{thebibliography}
\end{document}